Probing Berry curvature in magnetic topological insulators through resonant infrared magnetic circular dichroism.

Seul-Ki Bac,[1,2] Florian le Mardelé,[3] Jiashu Wang,[1] Mykhaylo Ozerov,[4] Kota Yoshimura,[1] Ivan Mohelský,[3] Xingdan Sun[3], Benjamin Piot,[3] Stefan Wimmer,[5] Andreas Ney,[5] Tatyana Orlova,[6] Maksym Zhukovskyi,[6] Günther Bauer,[5] Gunther Springholz,[5] Xinyu Liu,[1] Milan Orlita,[3,7] Kyungwha Park,[8] Yi-Ting Hsu,[1,*] Badih A. Assaf[1,**]

[1] Department of Physics and Astronomy, University of Notre Dame, Notre Dame IN, 46556, USA
[2] KU Leuven, Quantum Solid State Physics, Leuven, Belgium
[3] LNCMI-EMFL, CNRS UPR3228, Univ. Grenoble Alpes, Univ. Toulouse, Univ. Toulouse 3, INSA-T, Grenoble and Toulouse, France
[4] National High Magnetic Fields Laboratory, Florida State University, Tallahassee, Florida 32310, USA
[5] Institut für Halbleiter und Festkörperphysik, Johannes Kepler Universität, 4040 Linz, Austria
[6] Notre Dame Integrated Imaging Facility, University of Notre Dame, Notre Dame IN, 46556, USA
[7] Institute of Physics, Charles University, Ke Karlovu 5, Prague, 121 16 Czech Republic
[8] Department of Physics, Virginia Tech, Blacksburg, VA 24061, USA
*corresponding authors: yhsu2@nd.edu, bassaf@nd.edu

**Abstract:**

**Probing the quantum geometry and topology in condensed matter systems has relied heavily on static electronic transport experiments in magnetic fields. Yet, contact-free optical measurements have rarely been explored. Magnetic dichroism (MCD), the nonreciprocal absorption of circular polarized light, was theoretically linked to the quantized anomalous Hall effect in magnetic insulators and can identify the bands and momenta responsible for the underlying Berry Curvature (BC). Detecting BC through MCD faces two challenges: First, the relevant inter-band transitions usually generate MCD in the infrared (IR) range, requiring large samples with high quality. Second, while most magnetic materials are metallic, the relation between MCD and BC in metals remains unclear. Here, we report the observation of MCD in the IR range along with the anomalous Hall effect in thin film $MnBi_2Te_4$. Both phenomena emerge with a field-driven phase transition from an antiferromagnet to a canted ferromagnet. By theoretically relating the MCD to the anomalous Hall effect via BC in a metal, we show that this transition accompanies an abrupt onset of BC, signaling a topological phase transition from a topological insulator to a doped Chern insulator. Our density functional theory calculation suggests the MCD signal mainly originates from an optical transition at the Brillouin zone edge, hinting at a potential new source of BC away from the commonly considered Γ point. Our findings demonstrate a novel experimental approach for detecting BC and identifying the responsible bands and momenta, generally applicable to magnetic materials.**

Resonant magnetic circular dichroism (MCD) refers to the optical response of a magnetic material that preferentially absorbs light of a specific circular polarization direction (Fig. 1(b,c)).[1] Correspondingly, the absorption coefficient exhibits non-reciprocity with respect to the magnetization of the material. Resonant MCD has been traditionally employed to probe the Zeeman splitting and spin polarization of the conduction and valence bands of magnetic semiconductors and insulators [2–4]. Recent theoretical efforts in linking optical measurements to the quantum geometry of the Bloch states in magnetic and topological materials have brought new light to the information MCD carries [5–10]. In particular, MCD in magnetic insulators was found to be related to the Chern number[6] responsible for the DC quantized anomalous Hall effect and arising from Berry curvature (BC),[10] the imaginary part of the quantum geometric tensor[9]. MCD in the infrared (IR) frequency range, resonant with optical transitions involving BC-hosting bands is therefore a novel approach for detecting the band geometry and topology[6]. However, no experimental observations of a BC-induced IR-MCD has been reported.

Magnetic topological insulators characterized by a Chern number are, in fact, an ideal platform to experimentally detect BC through MCD. $MnBi_2Te_4$ is a recently discovered magnetic insulator of such kind, and was found to exhibit various topological phases in bulk and few-layer samples[11–16]. $MnBi_2Te_4$ crystallizes in a rhombohedral structure belonging to the space group R-3m, with a stacking septuple-layer sequence of Te-Bi-Te-Mn-Te-Bi-Te along the rhombohedral c-axis (Fig. 1(a)). Mn spins within the Mn layers are coupled ferromagnetically (intralayer coupling) whereas the interlayer coupling between septuple layers is antiferromagnetic. Such couplings lead to an A-type antiferromagnetic ground state that was found to be a topological insulator (TI) [17–20]. For thin films and flakes, recent experiments found a topological phase transition driven by an out-of-plane magnetic field from the antiferromagnetic TI state at zero field to a canted ferromagnetic state, and finally to ferromagnetic saturation.[17–19] The latter two phases with spontaneously broken time-reversal symmetry were suggested to be Chern insulators through the observation of chiral edge states[21–23]. As-grown thin films of $MnBi_2Te_4$ samples are however naturally metallic, making the detection of a resonant MCD challenging.[17,24]

The optical response of $MnBi_2Te_4$ and magnetic topological insulators in general has only recently become an attractive probe to unveil symmetry-broken and topological phases[5,8,25–27]. In particular, many magnetooptical responses become active and exhibit qualitative signatures upon the breaking of various symmetries[5,26–28]. For instance, thin films of $MnBi_2Te_4$ have been theoretically predicted to yield a large low energy Faraday and Kerr rotation when the ground state is ferromagnetic so that time-reversal symmetry is broken. According to Ref.[5], Kerr rotation is also expected in the antiferromagnetic state if inversion symmetry is broken. The origin of these predicted far- and mid-IR responses in $MnBi_2Te_4$ [5,27] is the large BC near the band edges, which generates a nonzero optical Hall conductivity $\sigma_{xy}(\omega)$. Experimentally, the magnetooptical response of $MnBi_2Te_4$ has never been studied in the IR range nor tied to the presence of BC. This is partly because large-area thin film samples with high quality are needed to accommodate the long wavelength in IR range, and partly because the relation between MCD and BC in the metallic state remains unclear.

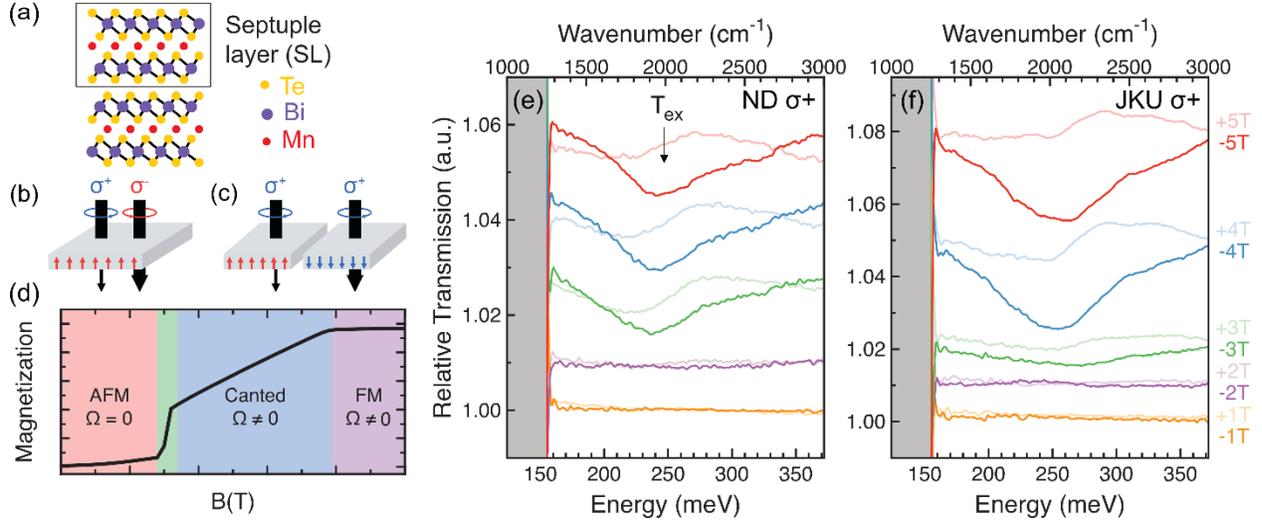

**Figure 1.** (a) Crystal structure of $MnBi_2Te_4$ consisting of septuple layers of Te-Bi-Te-Mn-Te-Bi-Te separated by a van der Waals gap. (b) Magnetic circular dichroism (MCD) manifesting as an absorption of light dependent on the direction of circular polarization. (c) MCD manifesting as a nonreciprocal absorption of circularly polarized light at opposite magnetic states of a material or opposite magnetic field. The red and blue arrows denote the direction of the magnetic moment. (d) Magnetic phase diagram of $MnBi_2Te_4$, where Ω is the Berry curvature. AFM: antiferromagnet, FM: ferromagnet. Relative transmission spectra acquired at different magnetic fields at T = 4.2 K with a defined polarization for the (e) ND and (f) JKU samples. $T_{ex}$ marks the optical transition studied in the remainder of this work.

In this work, we report a resonant infrared MCD in $MnBi_2Te_4$ thin films that onsets at the field-driven phase transition into the canted ferromagnetic state. The MCD and the associated optical transition are absent in the antiferromagnetic state at low field. We attribute the manifestation of the MCD to the abrupt onset of BC due to the topological phase transition from a time-reversal-symmetric antiferromagnetic TI to a doped Chern insulator (Fig. 1(c)) based on the following findings. First, we find that the MCD signal extracted from the measured optical absorption spectroscopy is tied to magnetization. Second, we theoretically show that the first negative moment of the antisymmetric (dissipative) part of the optical Hall conductivity $\sigma_{A,xy}(\omega)$ is related to the BC even in metallic states, and is proportional to the non-quantized anomalous Hall conductivity. This relation indicates that the onset of MCD signals originates from the onset of BC in the canted phase. Third, we experimentally show that the intensity of the measured MCD scales with the static anomalous Hall conductivity, further demonstrating that MCD is tied to the BC. Finally, a comparison between the magnetooptical response, the static Hall response, and first-principles calculations allows us to infer that this MCD likely originates from an optical transition at the vertical Brillouin zone boundary instead of at the commonly considered Γ point. Our findings directly detect the BC of Bloch bands along with the bands and momenta responsible for it through magnetooptical transitions, which enable future optical measurements aimed at probing quantum geometry and extracting topological indices.[6]

To unveil the MCD of $MnBi_2Te_4$, we employed magneto-optical infrared spectroscopy and measured thin films grown on GaAs(111) (ND) and $SrF_2$(111) (JKU) by two independent groups. With a fixed circular polarization, we detect the MCD as a nonreciprocity in the relative optical transmission when the polarity of the magnetic field is switched (Fig. 1(b)). The relative transmission is defined as $\tau=T_S(B)/T_S(0)$, where $T_s(B)$ is the transmission through the sample at magnetic field B parallel to the rhombohedral c-axis. The MCD is shown in Fig. 1(e,f). Below 3T, the relative transmission is close to 1 and has no energy dependence regardless of the direction of the magnetic field for either sample. At 3T and above, a pronounced minimum marked $T_{ex}$ in Fig. 1(e) is observed only when the magnetic field is negative. The intensity of the MCD is extracted at $|B|$ using:

$$\frac{\tau(+B) - \tau(-B)}{\tau(+B) + \tau(-B)} = MCD(\%)$$

The MCD is plotted as a function of energy in Fig. 2(a, b) and appears as a peak between 200 and 300 meV. The integrated intensity of the MCD signal, highlighted in Fig. 2(a,b), is subsequently extracted and shown in Fig. 2(c). The dichroism emerges at 3T and gains intensity with increasing field. In $MnBi_2Te_4$, 2T marks the phase boundary of the antiferromagnetic ground state. Above 2T, a surface spin-flop (SF) transition followed by a bulk spin-flop lead to a canted phase. The bulk spin-flop transition occurs at a magnetic field denoted by $B_{sf}$. It persists up to saturation to ferromagnetism at $B_{sat}$ [17–19]. For our sample, the anomalous Hall resistance shown in Fig. 2(d) elucidates these phase boundaries; in particular, $B_{sf}$ = 2.95T for ND and 3.1T for JKU, consistent with the delayed onset of the MCD observed in the JKU sample. (see Supplementary Figure 1 for complete Hall measurements and Supplementary Figure 2 for magnetization). The differences observed between samples are due to differences in growth conditions, which can include the Mn content, the substrate type and the thickness of the films. Magnetic canting breaks time-reversal symmetry, which induces a topological phase transition to a state with finite Berry curvature above the SF, activating the MCD (Fig. 2(c)). This onset is consistent with $MnBi_2Te_4$ transitioning progressively from an antiferromagnetic topological insulator state to a doped Chern insulator state above $B_{sf}$. The anomalous Hall resistance also increases in strength as of 2T, consistent with the activation of Berry curvature (see Fig. 2d). However, its value is not quantized since the bulk bands are partially filled for both samples.

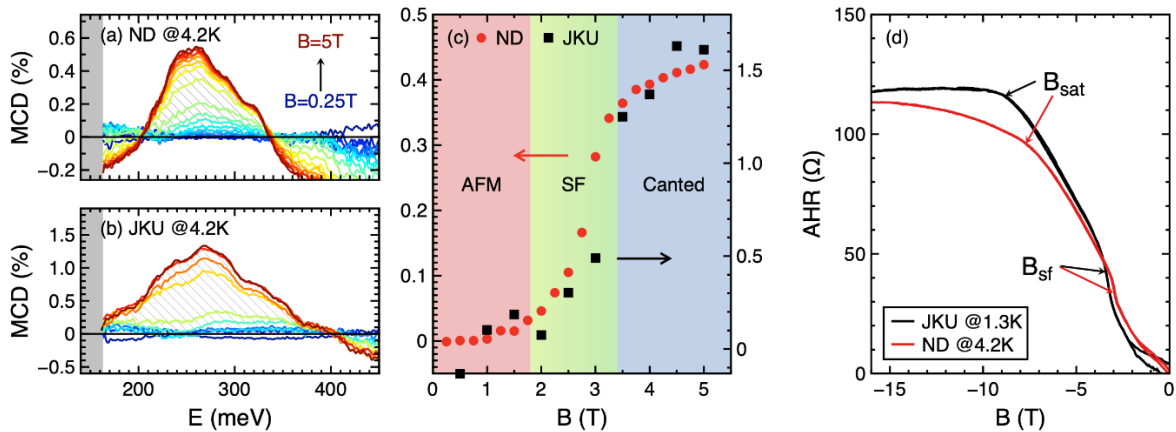

**Figure 2.** (a,b) Magnetic circular dichroism in % plotted as a function of energy for different magnetic field strengths (in steps of 0.25T) for the two $MnBi_2Te_4$ samples studied in Fig. 1, respectively (ND and JKU). MCD measurements are taken at T=4.2 K. (c) Integrated MCD versus magnetic field. AFM: Antiferromagnetic, SF: Spin-flop transition. (d) Anomalous Hall resistance (AHR) measured for the two samples. $B_{sf}$ denotes the spin-flop transition, and $B_{sat}$ denotes the saturation field.

We can understand the emergence of the MCD in the canted magnetic state through the following symmetry argument. The MCD signal can be expressed as the difference between the absorptions of opposite polarized lights $\sigma^+(\omega) - \sigma^-(\omega) = -2\text{Im}(\sigma_{H,xy}(\omega))$, where $\text{Im}(\sigma_{H,xy}(\omega))$ is the imaginary part of the dissipative (Hermitian) optical conductivity. $\text{Im}(\sigma_{H,xy}(\omega))$ is also given by the imaginary part of the antisymmetric optical conductivity $\sigma_{A,xy}(\omega) = \frac{1}{2}(\sigma_{xy}(\omega) - \sigma_{yx}(\omega))$. According to the Onsager relation, the optical conductivity obeys $\sigma_{xy}(\omega, M) = \sigma_{yx}(\omega, -M)$ for opposite magnetization $M$, and thus becomes symmetric in the Cartesian coordinate system when $M = 0$. It follows that the antisymmetric part vanishes at $M = 0$ such that $\sigma_{A,xy}(\omega, 0) = 0$. Therefore, we expect that the MCD signal vanishes in the antiferromagnetic state, grows with increasing magnetization strength $M$ upon entering the canted state, and saturates to a maximum value in the ferromagnetic state, as was observed in our experiment (see Fig. 2c).

While the MCD signal directly detects the magnetization strength, its first negative moment is known to be related to the Berry curvature and the static anomalous Hall conductivity $\sigma_{xy}(\omega = 0)$ through the Hall sum rule at temperature T=0K [6,10,29] as:

$$\text{Im}\left(\int_0^\infty d\omega \frac{\sigma_{A,xy}}{\omega}\right) = -\frac{\pi e^2}{2\hbar} \int \frac{d\mathbf{k}}{(2\pi)^3} \sum_n^{occ} f(\hbar \omega_{n,\mathbf{k}}) \Omega_{xy,n,\mathbf{k}} = \frac{\pi}{2}\sigma_{xy}(\omega = 0), \quad (1)$$

where -$e$ is the electron charge and, theoretically, the frequency (energy) cutoff $\omega_c \to \infty$. The key quantity in Eq. (1) is the Berry curvature $\Omega_{xy,n,\mathbf{k}}$ of band $n$

$$\Omega_{n,\mathbf{k},xy} = -2\hbar^2 \text{Im} \sum_m^{unocc} \frac{\langle u_{n,\mathbf{k}}|v_x|u_{m,\mathbf{k}}\rangle\langle u_{m,\mathbf{k}}|v_y|u_{n,\mathbf{k}}\rangle}{(\omega_{n,\mathbf{k}} - \omega_{m,\mathbf{k}})^2}$$

where $u_{n,\mathbf{k}}$ is the periodic part of the magnetic Bloch wavefunction, $\omega_{n,\mathbf{k}}$ is the energy of band $n$ at wavevector $k$, and $v_i$ is the velocity operator in the $i = x, y$ direction. Note that since the MnBi$_2$Te$_4$ samples we investigate are naturally doped, the number of occupied bands $n$ and unoccupied bands $m$ in the summations can be different at different crystal momenta $\mathbf{k}$. The right-hand side of the sum rule is proportional to the static anomalous Hall conductivity [30], which we experimentally found to onset in the canted ferromagnetic phase (see Fig. 2d). See Methods for the derivation of this sum rule. In our experiment, the frequency integration is over the infrared range that covers the low-energy regime in MnBi$_2$Te$_4$. We therefore expect that the sum rule approximately holds since contributions from bands far away from the Fermi level are suppressed by the frequency appearing in the denominator in the expression for $\Omega_{n,\mathbf{k},xy}$. Therefore, our measured MCD signal is directly tied to the BC and the observed anomalous Hall effect.

Moreover, through a spectroscopy of the optical transition exhibiting MCD, we show that its energy dependence is determined by the intrinsic magnetization of MnBi$_2$Te$_4$ and not by Zeeman splitting. The corresponding relative transmission spectra of the ND sample are presented in Fig. 3(a, b) for different magnetic field strengths. The data sets shown here were obtained using two different experimental setups at the Grenoble (GRE) and Tallahassee (TLH) magnet labs. Both measurements yield the fundamental optical transition labeled T$_{ex}$ between 230 and 260 meV. When measuring the optical transmission under circular polarization, T$_{ex}$ exhibits a dichroism up to 30T, as shown in Fig. 3(c), consistent with Fig. 1. Using the data shown in Fig. 3(a, b), we are able to track the field dependence of T$_{ex}$ by extracting the energy position of the minimum. Its transition energy increases between 3T and 8T and then saturates above 8T, as shown in Fig. 3(d). To understand this field dependence, we plot in Fig. 3(e) the magnetization of MnBi$_2$Te$_4$ computed using the Mills model of A-type antiferromagnets, with parameters constrained by previous experimental measurements.[17,18] By comparing Fig. 3(d) and Fig. 3(e), we conclude that the transition energy of T$_{ex}$ increases as the magnetization increases in the canted phase and then saturates once the system enters the ferromagnetic phase. Above 28T, the transition energy starts to increase again. This behavior is not captured by the Mills' model,[17,18,31] but is believed to originate from the response of Mn$_{Bi}$ antisites that couple antiferromagnetically to the Mn layers in MnBi$_2$Te$_4$.[32]

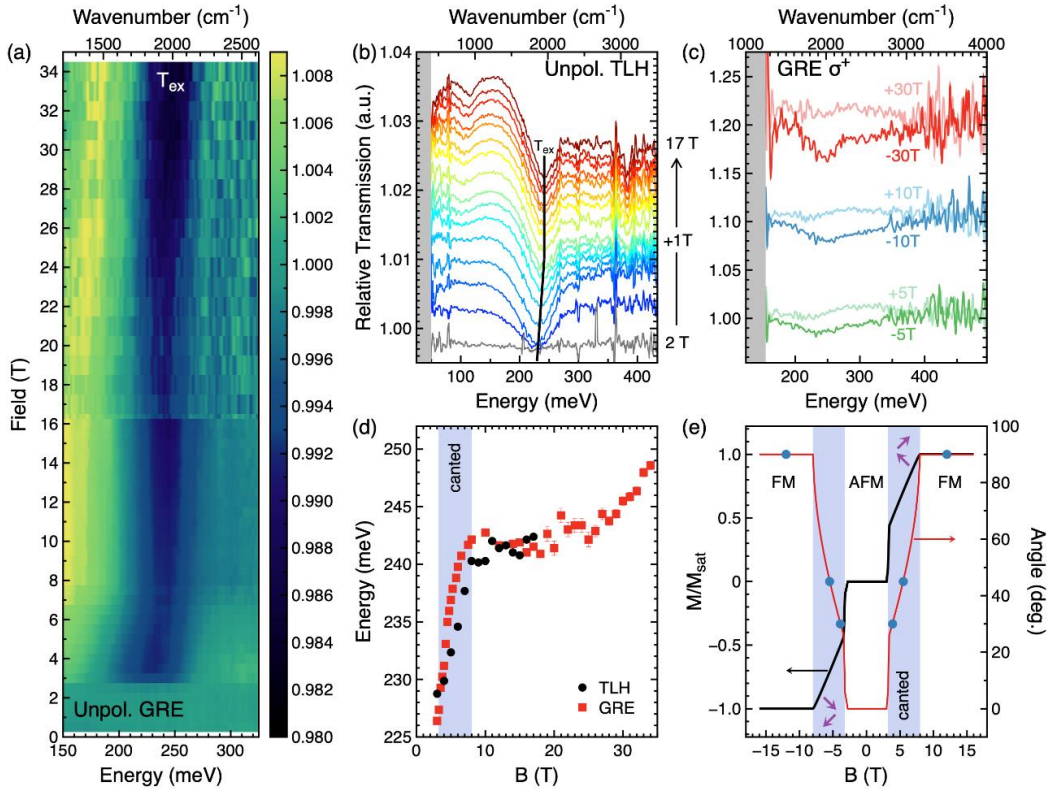

**Figure 3.** Unpolarized relative transmission spectra of sample ND acquired at different magnetic fields up to (a) 34T in Grenoble (GRE) and (b) 17T in Tallahassee (TLH) without a defined polarization. (c) Relative transmission spectra acquired with a defined σ$^+$ polarization up to 30T for sample ND. (d) Variation of the energy of T$_{ex}$ as a function of magnetic field measured on two different setups. The blue region highlights the canted magnetic phase of MnBi$_2$Te$_4$. (e) Magnetization relative to saturation (black curve) and canting angle with respect to the basal plane (red curve) of MnBi$_2$Te$_4$ computed using the Mills' model for a magnetic field applied along the c-axis. FM: Ferromagnet, AFM: Antiferromagnet. The spin orientation in the canted phase is illustrated by the purple arrows in (e). The blue circles represent the magnetic states used for the first-principles calculations.

The dichroic optical transition T$_{ex}$, along with the observed native doping of the samples studied here, raises the question of the specific band origin of this transition and of the MCD signal. We must confront the fact that samples with vastly different carrier densities (n or p) reproduce T$_{ex}$ at a comparable transition energy spanning only 30 meV. This is shown in Table 1. If the Fermi level E$_f$ crosses the conduction or valence band in a semiconductor, the absorption edge near the Γ-point is shifted to higher energies by a factor proportional to E$_f$; this phenomenon is referred to as the Moss–Burstein shift.

Essentially, samples with a higher carrier density would have an absorption edge at a higher energy. $T_{ex}$ does not follow this expectation, as shown in Table 1. We can thus conclude that $T_{ex}$ cannot be a transition at a small momentum off the Γ-point. Instead, $T_{ex}$ is nearly carrier density independent. This can imply that the magnetic exchange splitting at the Γ-point is giant and much larger than the differences in $E_f$ between different samples, which would cause $T_{ex}$ to occur between the conduction and valence band edges. Alternatively, $T_{ex}$ can be a transition occurring between other band extrema at higher energy or at a symmetry point other than Γ.

| Sample ID | Carrier density | Thickness/Substrate | Transition Energy ($T_{ex}$) at 5T |
|---|---|---|---|
| ND | n=2.2x10$^{19}$ cm$^{-3}$ | 41 nm/GaAs(111) | 234±2 meV |
| JKU | p=7.6x10$^{19}$ cm$^{-3}$ | 200 nm on SrF$_2$ | 257±1 meV |
| JKU-2 | n=2.3 x10$^{19}$ cm$^{-3}$ | 200 nm onBaF$_2$ | |
| ND-2 | n=4.6x10$^{19}$ cm$^{-3}$ | 27 nm/GaAs(111) | 237±6 meV |
| ND-3 | p=2.8 x10$^{19}$ cm$^{-3}$ | 83 nm/GaAs(111) | 255±4 meV |

**Table 1**. Carrier density, thickness and transition energy of the 4 samples studied here. The carrier density is extracted from Hall measurements at 5 K or below (see Supplementary Figure 1). The transition energy is extracted from Fig. 1 for ND and JKU and from Supplementary Fig. 3 for ND-2 and ND-3. The uncertainty is determined by fitting the peak minimum observed in the relative transmission. For sample ND, the uncertainty is the standard deviation taken on the two measurements shown in Fig. 3(d).

While our experimental data does not provide a definitive answer, density functional theory (DFT) calculations favor an interpretation involving a transition away from Γ at the Z-point of the BZ. We compare our experimental results to DFT calculations by computing the band structure for two canting angles (30° and 45°) as well as in the ferromagnetic state (see the blue dots in Fig. 3(e)). Two DFT schemes are compared in the ferromagnetic state, one that employs a single unit cell that is useful only for collinear magnetic states and the other employing a doubled unit cell that is used to model the magnetic texture in the canted state. Both are based on the scheme introduced in Ref.[33,34]. The calculation of the band structure for these three magnetic states allows us to track the field dependence of $T_{ex}$, excluding the cyclotron and Zeeman effects. The results are shown in Fig. 4(a-d). Here, $E_f$ is determined for sample ND from the Hall measurements shown in supplementary figure 1.

The interband transition at the Z-point is highlighted by the blue arrow and labeled A$_1$. It occurs between band extrema at the Brillouin zone edge. In this model, A$_1$ is not Pauli blocked, as shown in Fig. 4(c,d). The magnetization (and field) dependence of A$_1$ is extracted from Fig. 4(a-c) and plotted in Fig. 4(e). This field dependence shows remarkable agreement with the experimental data, with model parameters taken from ref. [35], despite a slight overestimation of the transition energies. DFT thus favors that a transition at the Z-point is the cause of $T_{ex}$ and its MCD. This DFT model also rules out that $T_{ex}$ occurs at the Γ-point because it does not produce the expected large magnetic exchange splitting that would lift one conduction band edge high enough in energy to allow its observation at a relatively fixed energy despite doping. However, further computational and experimental optical measurements of MnBi$_2$Te$_4$ are needed to confirm this phenomenon.

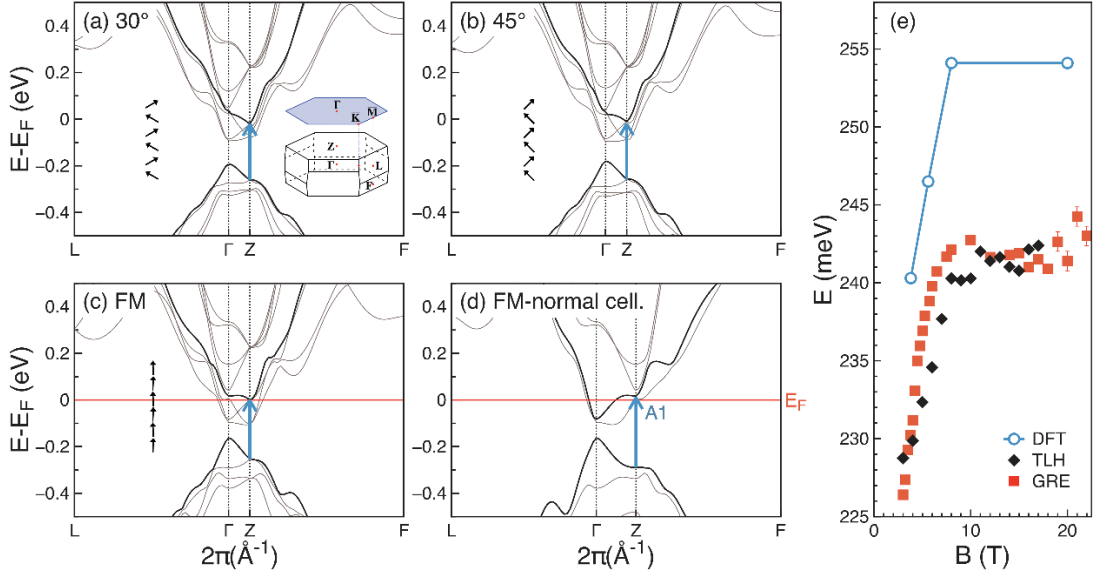

**Figure 4.** Density functional theory (DFT) calculations and their comparison with the experimental data. Band structure of MnBi$_2$Te$_4$ with canting angles of (a) 30° and (b) 45° with respect to the hexagonal basal plane of the films. Band structure of FM MnBi$_2$Te$_4$ computed using a single (c) and double (d) unit-cell model. $E_f$ marks the position of the Fermi energy determined by Hall effect measurements. A$_1$ – the blue arrow - highlights the transition from valence band 1 (VB1) to conduction band 2 (CB2) at the Z-point. The two bands of interest are bolded in (a-d). (e) Transition energy of A$_1$ (DFT) compared to the experimental data obtained in Grenoble (GRE) and Tallahassee (TLH) on MnBi$_2$Te$_4$ on GaAs. The magnetic field dependence of A$_1$ is obtained by converting the canting angle to magnetic field, as shown in Fig. 3(e).

We have thus shown that a Berry curvature is activated by the onset of the canted magnetic state in MnBi$_2$Te$_4$, which activates a resonant infrared MCD and a strong static anomalous Hall effect. The MCD is observed in samples grown under different conditions, and the optical transition at which the MCD occurs is observed in samples with vastly different carrier densities. A comparison with first-principles calculations indicates that this MCD is likely due to an optical transition occurring at the Z-point of MnBi$_2$Te$_4$. It may be possible that a giant magnetic exchange splitting lifts the conduction band edge above the Fermi energy for all the samples studied, resulting in a resonant MCD caused by the direct interband transition at the Γ-point. However, this giant splitting is not resolved by first-principles calculations of the bulk band structure.

We highlight here that a recent preprint[6] proposes MCD at energies resonant with the band edge interband transition as a novel measure of band topology and quantum geometry in ferromagnetic 3SL MnBi$_2$Te$_4$. Our results demonstrate how MCD can be used to track Berry curvature through the evolving magnetic phase diagram of MnBi$_2$Te$_4$ films under a magnetic field. They also motivate future studies of the resonant magnetooptical response of other magnetic materials expected to host a large Berry curvature, including Eu-based pnictides[36–38], ferromagnetic topological insulator MnSb$_2$Te$_4$[39], extrinsic magnetic topological insulators[40], chiral magnets[41] and altermagnets[42]. Broadly speaking, we demonstrate that resonant MCD and magnetooptical absorption can be regarded as interesting and complementary schemes to probe the quantum geometry tensor, since they are direct probes of the quantum mechanical transition dipole matrix element[8,43,44].

## Methods

**Derivation for the Hall Sum Rule.** The Hall Sum Rule in Eq. (1) can be obtained as follows. We will follow Ref. [43] closely, and the result is also consistent with more recent references[6,29]. The dissipative (Hermitian) part of the optical conductivity can be expressed using the Kubo–Greenwood formula as

$$\sigma_{H,xy} = -\frac{\pi e^2}{\hbar} \int \frac{d\mathbf{k}}{(2\pi)^3} \sum_n^{occ} \sum_m^{unocc} f_{nm,\mathbf{k}} \omega_{mn,\mathbf{k}} \langle u_{n,\mathbf{k}}|\partial_x u_{m,\mathbf{k}}\rangle\langle u_{m,\mathbf{k}}|\partial_y u_{n,\mathbf{k}}\rangle \delta(\omega - \omega_{mn,\mathbf{k}}),$$

where $\partial_i \equiv \partial_{k_i}$, $f_{nm,\mathbf{k}} = f(\hbar\omega_{n,\mathbf{k}}) - f(\hbar\omega_{m,\mathbf{k}})$ and $f(\hbar\omega_{n,\mathbf{k}})$ is the Fermi-Dirac distribution of band $n$ at momentum $\mathbf{k}$. The first negative moment of $\sigma_{H,xy}$ therefore has the form

$$\int_0^\infty d\omega \frac{\sigma_{H,xy}}{\omega}$$

$$= -\frac{\pi e^2}{\hbar} \int_0^\infty d\omega \int \frac{d\mathbf{k}}{(2\pi)^3} \sum_n^{occ} \sum_m^{unocc} f_{nm,\mathbf{k}} \frac{\omega_{mn,\mathbf{k}}}{\omega} \langle u_{n,\mathbf{k}}|\partial_x u_{m,\mathbf{k}}\rangle\langle u_{m,\mathbf{k}}|\partial_y u_{n,\mathbf{k}}\rangle \delta(\omega - \omega_{mn,\mathbf{k}})$$

$$= -\frac{\pi e^2}{\hbar} \int \frac{d\mathbf{k}}{(2\pi)^3} \sum_n^{occ} \sum_m^{unocc} f_{nm,\mathbf{k}} \langle u_{n,\mathbf{k}}|\partial_x u_{m,\mathbf{k}}\rangle\langle u_{m,\mathbf{k}}|\partial_y u_{n,\mathbf{k}}\rangle$$

$$= \frac{\pi e^2}{\hbar} \int \frac{d\mathbf{k}}{(2\pi)^3} \sum_n^{occ} \sum_m^{unocc} f_{nm,\mathbf{k}} \langle \partial_x u_{n,\mathbf{k}}|u_{m,\mathbf{k}}\rangle\langle u_{m,\mathbf{k}}|\partial_y u_{n,\mathbf{k}}\rangle.$$

Note that for the metallic case we have, the number of occupied (unoccupied) states $n$ ($m$) included in the summation is momentum $\mathbf{k}$-dependent. At T=0 K, we can simplify the expression by setting $f_{nm,\mathbf{k}} = 1$ and denoting the momentum dependence in the summations by changing $\sum_n^{occ}, \sum_m^{unocc} \to \sum_n^{occ\prime}, \sum_m^{unocc\prime}$. Moreover, the Berry curvature $\Omega_{n,\mathbf{k},xy}$ in Eq. (2) can be expressed in terms of the imaginary part of the quantum metric tensor $\Omega_{n,\mathbf{k},xy} = -2\,\mathrm{Im}\sum_m^{unocc\prime} \langle \partial_x u_{n,\mathbf{k}}|u_{m,\mathbf{k}}\rangle\langle u_{m,\mathbf{k}}|\partial_y u_{n,\mathbf{k}}\rangle$ at T=0 K, where we have used the relation $\langle u_{n,\mathbf{k}}|v_i|u_{m,\mathbf{k}}\rangle = \frac{\omega_{nm}}{\hbar}\langle u_{n,\mathbf{k}}|\partial_i u_{m,\mathbf{k}}\rangle$ with $\partial_i = \partial_{k_i}$. We therefore arrive at the Hall sum rule in Eq. (1) by taking the imaginary part on both sides

$$\mathrm{Im}\int_0^\infty d\omega \frac{\sigma_{H,xy}}{\omega} = -\frac{\pi e^2}{2\hbar} \int \frac{d\mathbf{k}}{(2\pi)^3} \sum_n^{occ\prime} \Omega_{n,\mathbf{k},xy}.$$

Note that the sum rule is applicable to metallic states since we allow the numbers of occupied and unoccupied states in the summations to be different at each momentum. In this case, the right-hand side becomes the well-known anomalous dc Hall conductivity $\sigma_{xy}(\omega = 0)$ at T=0 [30] for multiband systems, multiplied by $\frac{\pi}{2}$. We therefore arrive at Eq. (1).

**MBE growth – Notre Dame.** $MnBi_2Te_4$ is grown on GaAs (111) by molecular beam epitaxy. Prior to growth, the native oxide of GaAs is desorbed, and a buffer layer of GaAs is grown to obtain a pristine surface in a III-V MBE chamber. The substrates are then transferred to a group VI chamber in situ without breaking vacuum and are exposed to a flux of Te for 20 minutes. A buffer layer of $Bi_2Te_3$ is grown, followed by $MnBi_2Te_4$. $MnBi_2Te_4$ is grown using a sequential deposition of Bi-Te, Mn-Bi-Te and Te repeated 10-40 times at a substrate temperature of 370°C depending on the desired thickness. The resulting film (ND) 41nm thick. The thicknesses of ND-2 and ND-3 are shown in Table 1.

**MBE growth – Johannes Kepler University.** Epitaxial $MnBi_2Te_4$ layers were grown by molecular beam epitaxy on cleaved (111) $SrF_2$ and $BaF_2$ substrates in a Varian Gen II system by co-deposition of $Bi_2Te_3$ and Mn under excess Te flux. The $Mn/Bi_2Te_3$ flux ratio measured by a quartz microbalance was adjusted to obtain the $MnBi_2Te_4$ phase, and the layer thickness was 200 nm. Deposition was carried out at a growth temperature of 360°C, measured by an infrared pyrometer. In situ high-energy electron diffraction (RHEED) was employed to verify the conditions for two-dimensional growth.

**Structural Characterization.** Standard X-ray diffraction measurements are performed using a Bruker D8 Discover diffractometer and Malvern Aeris powder diffractometer both equipped with a Cu-Kα-source. (See Data in Supplementary Note 4 and Supplementary Figure 5).

**Magnetic Characterization.** Magnetic properties were determined by magnetization measurements M(H) (see Supplementary Figure 2) at 2 K and 300 K with applied in-plane and out-of-plane external magnetic field. The measurement employs a superconducting quantum interference device (SQUID) magnetometer (Quantum Design MPMS-XL).

**Transmission electron microscopy**. High-resolution cross-sectional TEM images were acquired using a double tilt holder and probe-corrected Spectra 30-300 transmission electron microscope (Thermo Fisher Scientific, USA) equipped with a field emission gun operated at 300 kV. STEM images were acquired using a Panther STEM detector (Thermo Fisher Scientific, USA) in high-angle, annular dark field (HAADF) mode and bright field (BF) mode. For compositional analysis, energy-dispersive X-ray spectroscopy (EDX) maps were obtained in STEM mode using a Super-X EDX system (Thermo Fisher Scientific, USA) equipped with 4 windowless silicon drift detectors. TEM samples were prepared by focused ion beam etching using the standard lift-out technique. (See TEM in Supplementary Figure 6).

**Magnetooptical measurements – Tallahassee.** Magneto-infrared experiments were performed using a commercial Fourier transform infrared spectrometer coupled with a vertical bore superconducting magnet reaching 17.5 T at the National High Magnetic Fields Lab. The IR radiation propagates inside an evacuated metal tube from the spectrometer to the top of the magnet, whereas a brass light pipe is used to guide the IR radiation down to the sample space. A parabolic cone was used to collect the IR beam at the sample, while two mirrors reflected the beam back up toward the Si composite bolometer mounted just a short distance above the sample space. The sample is cooled down to 5 K using low pressure He exchange gas. The IR spectra were recorded in the mid-IR and far-IR at a fixed magnetic field, while the field was stepped between 0 and 17.5T with increments of 1T. All measurements are carried out in the Faraday geometry with the incident beam propagating perpendicular to the surface and along the direction of the applied field. The reported spectra are obtained by dividing the signal at a finite magnetic field by the zero-field spectrum.

**Magnetooptical measurements – Grenoble.** The magneto-optical response of $MnBi_2Te_4$ was probed in transmission mode and in the Faraday configuration with the magnetic field perpendicular to the sample surface. The radiation from a globar was delivered to the sample via light-pipe optics and was analyzed by a Bruker Vertex 80v Fourier transform spectrometer. In both experiments, employing either a superconducting coil (from 0 to 16 T) or a resistive coil (from 0 to 34 T), radiation was detected by a composite bolometer positioned directly beneath the sample. In experiments performed with circularly polarized light, a glass linear polarizer and a zero-order MgF$_2$ quarter-wave plate (centered at λ = 5 μm) were used.

**Magnetotransport measurements.** Magnetotransport measurements at ND are carried out in a 16T/1.5K Oxford Instruments system coupled to a DC current source and standard nanovoltmeters. Five Indium contacts are soldered on a rectangular piece of the sample. An excitation current of 100μA is used for these measurements. Magnetotransport measurements at Grenoble on sample JKU measurements are carried out down to T=1.4K in a variable temperature insert fitted into a 16T superconducting magnet. Electrical contacts (either indium or silver-based) were deposited on rectangular-shaped samples in a "Hall bar-like" geometry. Four-terminal measurements were performed either with DC or with low-frequency lock-in techniques, with currents varying between 100 nA and 100 μA.

**Density Functional Theory.** We performed density functional theory (DFT) calculations for bulk MnBi$_2$Te$_4$ including spin-orbit coupling and the on-site Coulomb repulsion term U=5.34 eV, using the VASP code [45,46]. The value of U was chosen following Ref. [35] to take into account the strong correlation from Mn d-orbitals. We used projector-augmented-wave (PAW) pseudopotentials [47,48] and the Perdew–Burke–Ernzerhof (PBE) [49] generalized gradient approximation for the exchange-correlation functional. We also used the experimental lattice constants [50] for the MnBi$_2$Te$_4$ structure without further geometry relaxation. We considered the ferromagnetic structure where the magnetic moments of the Mn sites are all oriented along the z (c-) axis, i.e., the [111] direction (in the rhombohedral lattice), as well as several other magnetic structures where the magnetic moments of the Mn sites are tilted from the *z*-axis in the xz plane (see Fig. 1), following the results in Ref.[17]. Here, the *x*-axis is along the [1$\bar{1}$0] direction in the rhombohedral lattice. We considered tilt angles θ of 10°, 30°, and 45° from the z-axis. In the tilted magnetic structures, the unit cell was doubled to incorporate the magnetic order. Due to the doubled unit cell, a double degeneracy appears in the bands at the Z point. For the double unit cell, we sampled 9 × 9 × 9 k-points, while for the normal unit cell (consisting of seven atoms), we sampled 18 × 18 × 18 k-points. A kinetic cutoff energy of 270 eV was used. In addition, van der Waals interactions were included using the DFT-D3 method of Grimme with a zero-damping function [51].

**Data availability.** The data that support the findings of this study are available from the corresponding authors upon reasonable request.

**Acknowledgments.** BAA, JW, KY, SKB and XL acknowledge support from National Science Foundation grant DMR-1905277. KY was partly supported by Department of Energy Basic Energy Science Award DE-SC0024291. Y.-T.H. acknowledges support from National Science Foundation grant DMR-2238748. A portion of this work was performed at the National High Magnetic Field Laboratory, which is supported by National Science Foundation Cooperative Agreements No. DMR-1644779, DMR-2128556, and the State of Florida. GS, SW, AN, and GB acknowledge support from the Austrian Science Funds grant I-4493, AI0656811/21, as well as the JKU-Linz grant LIT-2022-11-SEE-131. XS and BP are supported by ANR-20-CE30-0015-01.

**Author contributions.** SKB, FLM, IM and JW carried out magnetooptical measurements with assistance from KY, Orlita and Ozerov. SKB carried out magnetotransport measurements and X-ray diffraction on the ND samples. BP and XS carried out magnetotransport measurements on the JKU sample. YTH carried out the symmetry analysis. KP carried the first principles calculations. TO and MZ carried out TEM measurements. XL and GS grew the ND and JKU samples, respectively. AN and SW contributed to the characterization of the JKU samples. BAA, YTH, KP, SKB, FLM and Orlita interpreted the data and confronted the experiments and the theory. YTH, SKB and BAA wrote the manuscript with input from KP, Orlita, GB and GS.